# Metallic-Phase-$Fe_3GaTe_2$ Enabled Interface Engineering for Self-Powered and High-Gain $WS_2$ Photodetectors

*Wajid Ali[a,b†], Ming Huang[b,c†], Juan Li[b], Jianhua Huang[b], Liuli Yang[b], Sajid Ur Rehman[d], Chinmay K. Mohanty[a], Zahir Muhammad[e], Ziwei Li[b,*] and Maciej R. Molas[a,*]*

[a] Faculty of Physics, University of Warsaw, Warsaw 02-093, Poland.

[b] Hunan Institute of Optoelectronic Integration, College of Materials Science and Engineering, Hunan University, Changsha 410082, P. R. China

[c] College of Physics and Electronic Engineering, Xinyang Normal University, Xinyang 464000, P. R. China

[d] School of Science, Minzu University of China, Beijing 100081, P. R. China

[e] National Key Laboratory of Spintronics, Hangzhou International Innovation Institute, Beihang University, Hangzhou 311115, P.R. China

Corresponding authors: *E-mail: ziwei_li@hnu.edu.cn, maciej.molas@fuw.edu.pl

**Abstract**

Two-dimensional transition-metal dichalcogenides offer strong light-matter interaction but suffer from inefficient carrier separation and contact-related losses in photodetectors. Here, we demonstrate a high-gain $WS_2/Fe_3GaTe_2$ van der Waals heterostructure photodetector, where metallic $Fe_3GaTe_2$ serves as an active interfacial contact. The work-function mismatch, together with interfacial charge redistribution and asymmetric contact geometry, contributes to a built-in field that supports self-powered photodetection at zero bias. Under 450 nm illumination, the device delivers a zero-bias responsivity of 23.5 A/W and an apparent external quantum efficiency of $6.4\times 10^3$%. At –1 V biasing, the heterostructure exhibits photoresponse at 450, 520 and 633 nm, achieving a responsivity of $9.7 \times 10^3$ A/W and a noise derived specific detectivity of $2.3 \times 10^{13}$ Jones at 100 Hz under 450 nm illumination. The high photoresponse is attributed to interfacial carrier separation, efficient extraction, and a likely contribution from trap-assisted photogating in multilayer $WS_2$. These results establish $Fe_3GaTe_2$-enabled interface engineering as an effective route for self-powered, highly sensitive 2D photodetectors.

## 1.Introduction

Two-dimensional transition-metal dichalcogenides (TMDCs) have emerged as promising materials for next-generation photodetectors because of their strong light–matter interaction, atomically thin geometry, and compatibility with van der Waals (vdW) integration [1–10]. Among them, $WS_2$ is particularly attractive owing to its visible-range optical absorption, thickness-dependent electronic structure, and stable photoresponse at room temperature [11,12]. However, the practical performance of $WS_2$-based photodetectors is still strongly affected by inefficient photocarrier separation, interfacial recombination, and contact-related losses, which often require an external bias to obtain high photoresponse [13,14]. Interface engineering provides an effective route to overcome these limitations. In vdW heterostructures (HS), band alignment and interfacial charge transfer can be tailored without the constraints of lattice matching, enabling the formation of built-in electric fields for efficient photocarrier separation and self-powered photodetection. At the same time, low-resistance contacts are essential for achieving high responsivity and detectivity.

Several contact-engineering strategies have been explored to improve carrier transport in two-dimensional optoelectronic devices. Graphene and black phosphorus electrodes can improve interfacial carrier transport, but they are often limited by high dark current and environmental instability, respectively [15,16]. Conventional metal contacts such as Au, Ag, and Ti can form Schottky barriers at $WS_2$ interfaces, which impede carrier injection, increase contact resistance, and suppress photocurrent generation [17–19]. Low-work-function metals such as Sc and Ti can reduce barrier heights in TMDC devices, but oxidation and interfacial diffusion may compromise long-term stability [20,21]. Metallic layered materials such as $NbSe_2$ and $TaS_2$ provide cleaner vdW interfaces, although nonideal work-function matching can still limit band alignment and carrier extraction [22,23]. In this context, $Fe_3GaTe_2$ (FGT), a metallic vdW ferromagnet with high electrical conductivity and a favorable work-function relationship with $WS_2$, offers a promising interfacial contact for efficient charge separation and extraction [17,24]. Compared with conventional evaporated metals and graphene electrodes, FGT can form a cleaner $WS_2$ interface, potentially reducing contact-related losses and supporting built-in-field-assisted photodetection [25,26]. However, FGT-enabled interface engineering for self-powered and high-gain $WS_2$ photodetection remains largely unexplored.

In this work, we demonstrate a self-powered photodetector based on a $WS_2$/FGT vdW metal–semiconductor HS that enables efficient interfacial charge separation and zero-bias photoresponse. Owing to favorable band alignment at the $WS_2$/FGT interface and asymmetric contact-induced internal electric fields, the device exhibits a high self-powered responsivity of 23.5 A/W under 450 nm illumination. The apparent external quantum efficiency exceeding 100% is consistent with an internal gain mechanism, likely involving trap-assisted photogating in multilayer $WS_2$. The photodetector further maintains zero-bias visible-wavelength operation, with responsivities of 19.8 A/W at 520 nm and 6.1 A/W at 633 nm. These results demonstrate that $WS_2$/FGT vdW HS provide an effective platform for realizing low-power and high-sensitivity photodetection based on engineered metal–semiconductor interfaces.

## 2.Results and Discussion

Figure 1a schematically illustrates the device configuration, in which multilayer $WS_2$ partially overlaps with metallic FGT while the two materials are connected to asymmetric Au electrodes. The optical microscopy image in Figure 1b confirms the spatial overlap between $WS_2$ and FGT, while the corresponding atomic force microscopy (AFM) analysis in Figure 1c reveals well-defined flake thickness and a clean interface. Detailed structural characterization of the individual FGT and $WS_2$ crystals, together with the HS height profile, is provided in Supporting Information Figures S1 and S2.

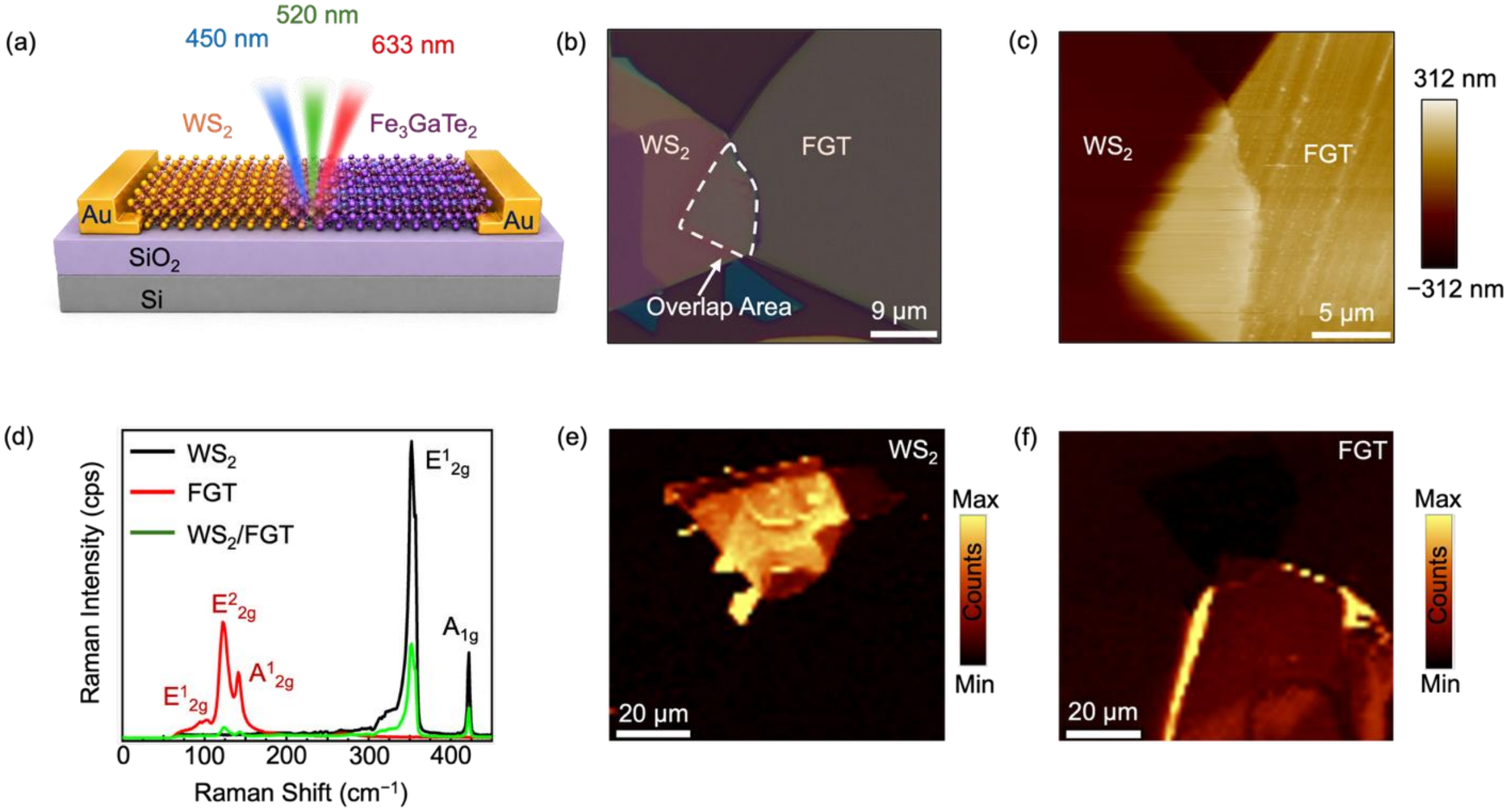

Fig. 1 (a) Schematic of $WS_2$/FGT vdW HS. (b) Optical image of the device. (c) AFM map of $WS_2$/FGT HS. (d) Raman spectrum acquired from the HS region, showing the characteristic vibrational modes of $WS_2$ and FGT. (e, f) Raman intensity maps of the characteristic $WS_2$ and FGT modes, respectively, confirming the spatial distribution and overlap of the two materials.

Raman spectroscopy observation in Figure 1d illustrates characteristic vibrational modes corresponding to both FGT and $WS_2$. The in-plane ($E^1_{2g}$, $E^2_{2g}$) and out-plane ($A^1_{2g}$) FGT modes at 101, 122.8, and 141.5 cm$^{-1}$, respectively are consistent with previously reported observation [27]. Similarly, the well-known $WS_2$ ($E_{2g}$ =351.2 cm$^{-1}$) and ($A_{1g}$ =422.1 cm$^{-1}$) modes shows a slight shift after integration with FGT, indicating interfacial interaction between the two layers that may arise from charge transfer and vdW coupling effects [28,29].

The integrated intensity maps (Figure 1e and f) reveal spatial distribution of the two materials and clearly identify the overlapped HS region. The localized reduction in $WS_2$ Raman intensity within the HS overlapped region indicates that the metallic FGT layer alters the local optical and electronic environment of the overlying $WS_2$ layer. This attenuation can be associated with interfacial charge transfer and enhanced dielectric/electrostatic screening from the metallic FGT layer [17]. Consistently, the photoluminescence (PL) intensity of $WS_2$ is also quenched after contact with FGT, as shown in Supporting Information Figure S3, suggesting enhanced nonradiative relaxation and efficient interfacial carrier transfer. These spectroscopic results confirm the formation of a coupled $WS_2$/FGT vdW interface, which provides the structural basis for photodetection discussed below.

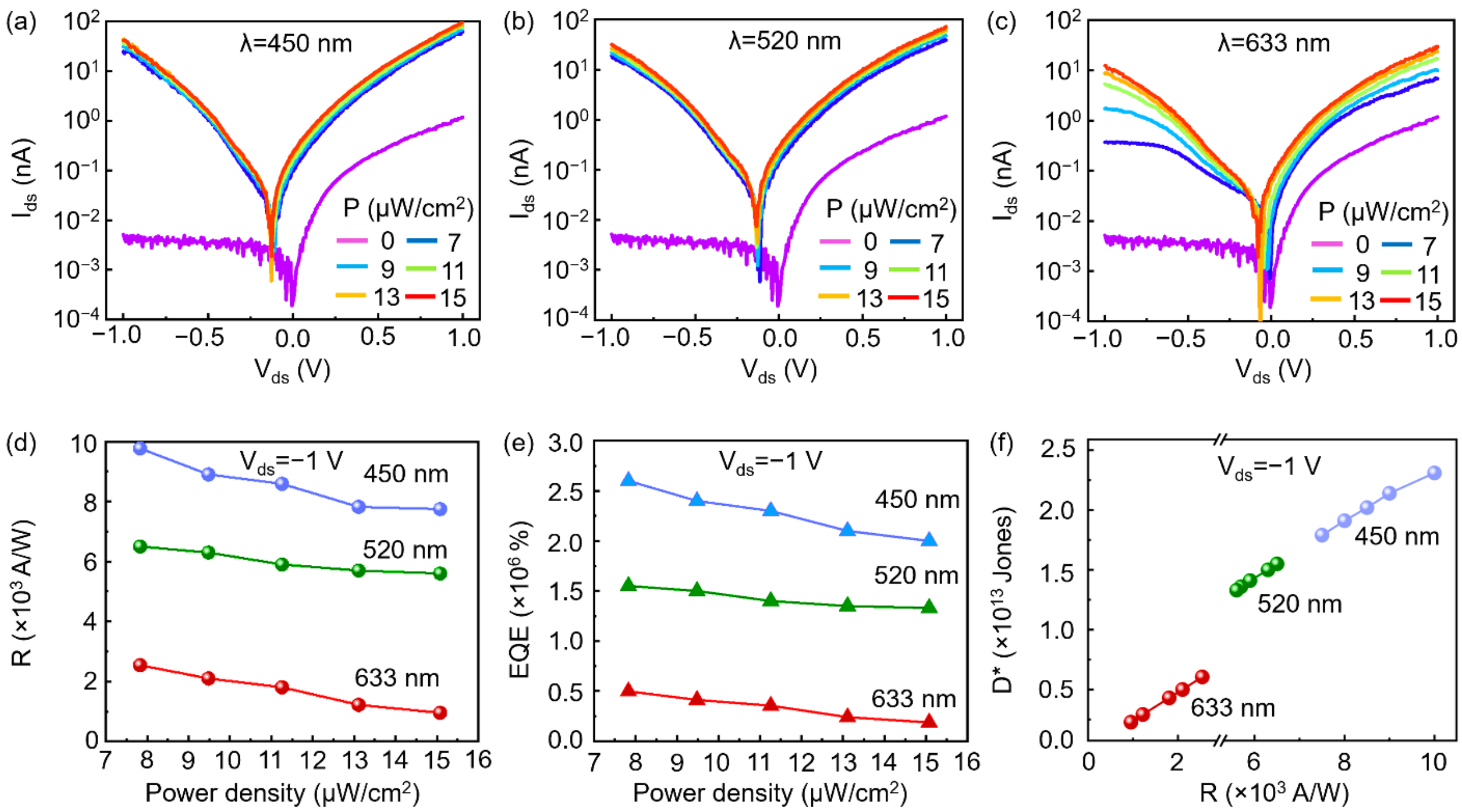

Fig. 2 (a–c) *I–V* characteristics measured under dark conditions and under 450 nm, 520 nm, and 633 nm laser illumination, respectively. Incident light with power density *P* ranges from 0 to 15 $\mu$W cm$^{-2}$. (d,e) Responsivity and external quantum efficiency at $V_{ds}$= −1 V. (f) Noise-derived specific detectivity calculated at 100 Hz and $V_{ds}$= −1 V using the experimentally measured current-noise amplitude spectral density. For all measurements $V_g$= 0 V.

Photoresponse of the $WS_2$/FGT HS was systematically investigated by current–voltage (I–V) measurements under controlled illumination at room temperature, as shown in Figure 2. The drain–source current ($I_{ds}$) was recorded as a function of drain–source voltage ($V_{ds}$) from −1 to +1 V under dark conditions and under 450, 520, and 633 nm laser illumination, with grounded gate ($V_g$ = 0 V). As shown in Figures 2a–c, the device exhibits pronounced asymmetric transport in the dark, with a rectification factor of 285.39 at $|V_{ds}|$ = 1 V, indicating the formation of an asymmetric Schottky-type junction [30]. This rectifying behavior arises from the $WS_2$/FGT metal–semiconductor interface together with the asymmetric Au/$WS_2$/FGT contact geometry. The lower work function of FGT relative to $WS_2$ favors interfacial charge redistribution and band bending, generating a built-in electric field that assists photocarrier separation. Under illumination, the current increases significantly for all excitation wavelengths, confirming efficient photogeneration and carrier extraction in the HS.

The photoresponsivity "R" of the device, defined as the electrical output per unit incident optical power, was calculated using the relation [31].

$$R = \frac{I_{ph}}{P_{light} \times A} \quad (1)$$

Here $I_{ph} = I_{light} - I_{dark}$ is the net photocurrent, $P_{light}$ denotes the incident optical power density, and $A$=80 $\mu m^2$ is the effective photosensitive area of the $WS_2$/FGT heterostructure. At −1 V, the device achieves high responsivities of $9.7\times10^3$, $6.5\times10^3$, and $2.5\times10^3$ A/W under 450, 520, and 633 nm illumination, respectively (Figure 2d). The stronger response at shorter wavelength is consistent with enhanced optical absorption and more efficient photocarrier generation in multilayer $WS_2$. For all wavelengths, R decreases with increasing incident power, reflecting trap-state saturation and enhanced recombination at higher carrier densities. This sublinear behavior (also in Figure S4) is consistent with trap filling and suggests a contribution from photogating to the enhanced photoresponse [32,33].

External quantum efficiency (EQE) was calculated as [34]

$$\text{EQE} = \frac{hc}{q} \times \frac{R}{\lambda} \times 100\% \quad (2)$$

Where h, c, q, and λ are Planck's constant, the speed of light, the elementary charge, and the excitation wavelength, respectively. As shown in Figure 2e, the apparent EQE follows the same power-dependent trend as R and reaches a maximum value of $2.6\times10^6$ % under 450 nm illumination at $V_{ds}$ = −1 V. Although an apparent EQE exceeding 100% suggests an internal amplification process, EQE and photoconductive gain are conceptually distinct, and EQE alone does not provide a direct quantitative measurement of gain. Together with the sublinear power dependence, this result suggests a possible contribution from trap-assisted carrier dynamics in multilayer $WS_2$.[35]

The noise-derived specific detectivity was calculated using the experimentally measured current-noise amplitude spectral density:[35]

$$D^*(f) = \frac{R\sqrt{A}}{I_n(f)} \quad (3)$$

Where *R* is the responsivity, *A* is the effective photosensitive area, and $I_n(f)$ is the measured current-noise amplitude spectral density. At $V_{ds}$ = −1 V and 100 Hz, $I_n$ = $3.75\times10^{-13}$ A Hz $^{-1/2}$. Using *A*=80 *μ*m², the maximum noise derived specific detectivities are $2.3\times10^{13}$, $1.5\times10^{13}$, $5.9\times10^{12}$ Jones under 450, 520, and 633 nm illumination, respectively. Further details are provided in Supporting Information Figure S6. Motivated by the illumination-induced shift of the current-null point observed in Figures 2a–c, we further evaluated the device performance under zero-bias conditions. As shown in Figure 3, the device maintains a pronounced power-density-dependent photoresponse at $V_{ds}$ = 0 V, confirming that photocurrent generation does not require an externally applied voltage. The maximum zero-bias responsivities reach 23.5, 19.8, and 6.1 A/W under 450, 520, and 633 nm illumination, respectively (Figure 3a). The corresponding apparent EQE values are approximately $6.4\times 10^3$%, $4.7\times 10^3$% and $1.2\times 10^3$%, respectively (Figure 3b). Both responsivity and EQE decrease with increasing incident power density, consistent with progressive trap filling and enhanced carrier recombination at higher photocarrier densities. Although these observations are consistent with internal-field-assisted self-powered operation, possible photothermoelectric and trap-assisted contributions cannot be completely excluded.

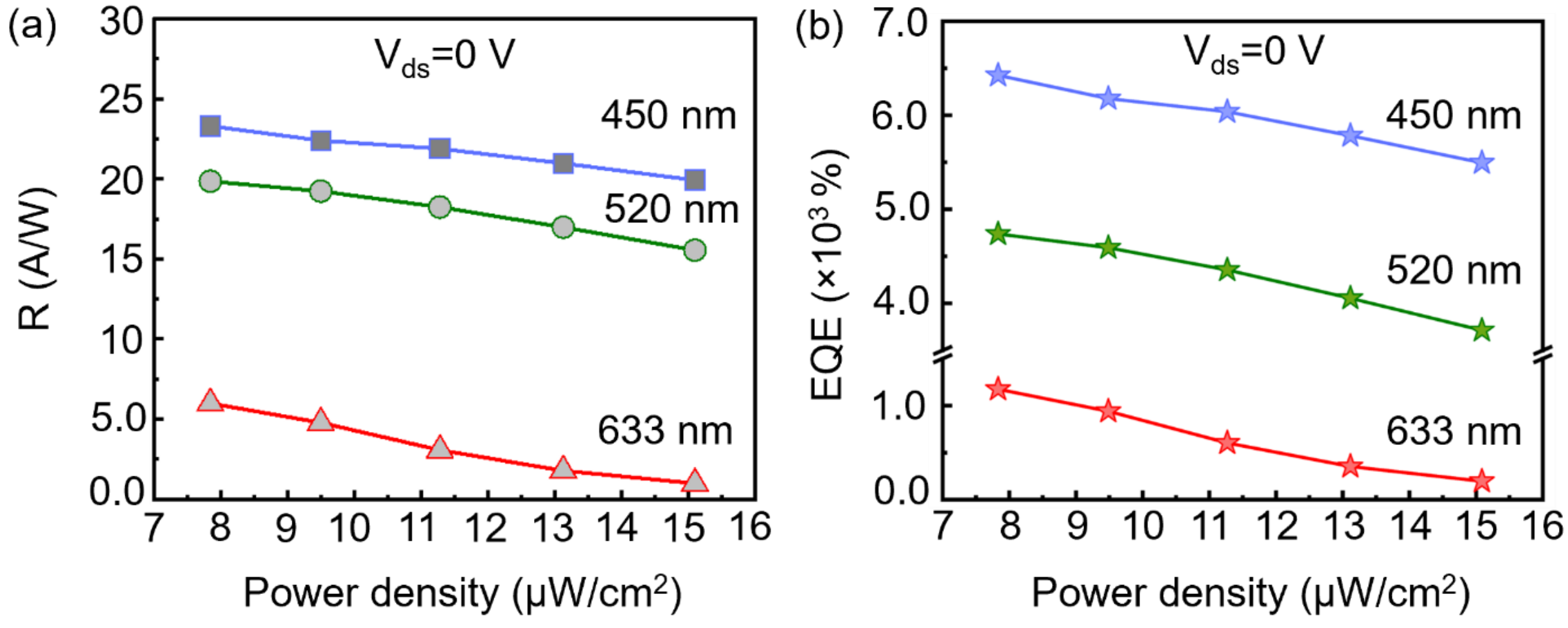

Fig. 3 Power-density-dependent (a) responsivity and (b) external quantum efficiency of the device at $V_{ds}$= 0 V under 450, 520 and 633 nm illumination. For all measurements, $V_g$= 0 V.

To understand the carrier dynamics and gain mechanism, incident-power dependence of the photocurrent was analyzed using the power-law relation $I_{ph} \propto P^{\alpha}$. As shown in Figure 4a, the photocurrent exhibits sublinear behavior at $V_{ds}$ = −1 V, with extracted α values of 0.60, 0.77, and 0.89 for 450, 520, and 633 nm excitation, respectively. The deviation from linearity is consistent with trap-assisted carrier dynamics and recombination processes in multilayer $WS_2$. The stronger sublinearity at 450 nm is consistent with more efficient photocarrier generation, faster trap filling, and partial screening of the built-in field at higher carrier density. In contrast, the larger α value at 633 nm indicates weaker trap saturation because of the lower photogeneration rate. The operational stability of the device was evaluated under periodically modulated illumination, as shown in Figure 4b. The device exhibits stable and reproducible switching behavior for all three wavelengths, with on/off ratios of approximately 1.1 × $10^4$, 7.9 × $10^3$, and 2.8 × $10^3$ for 450, 520, and 633 nm, respectively. This wavelength-dependent trend agrees with the R and D results, confirming that the photoresponse is governed mainly by optical absorption in multilayer $WS_2$ and efficient carrier separation at the $WS_2$/FGT interface.

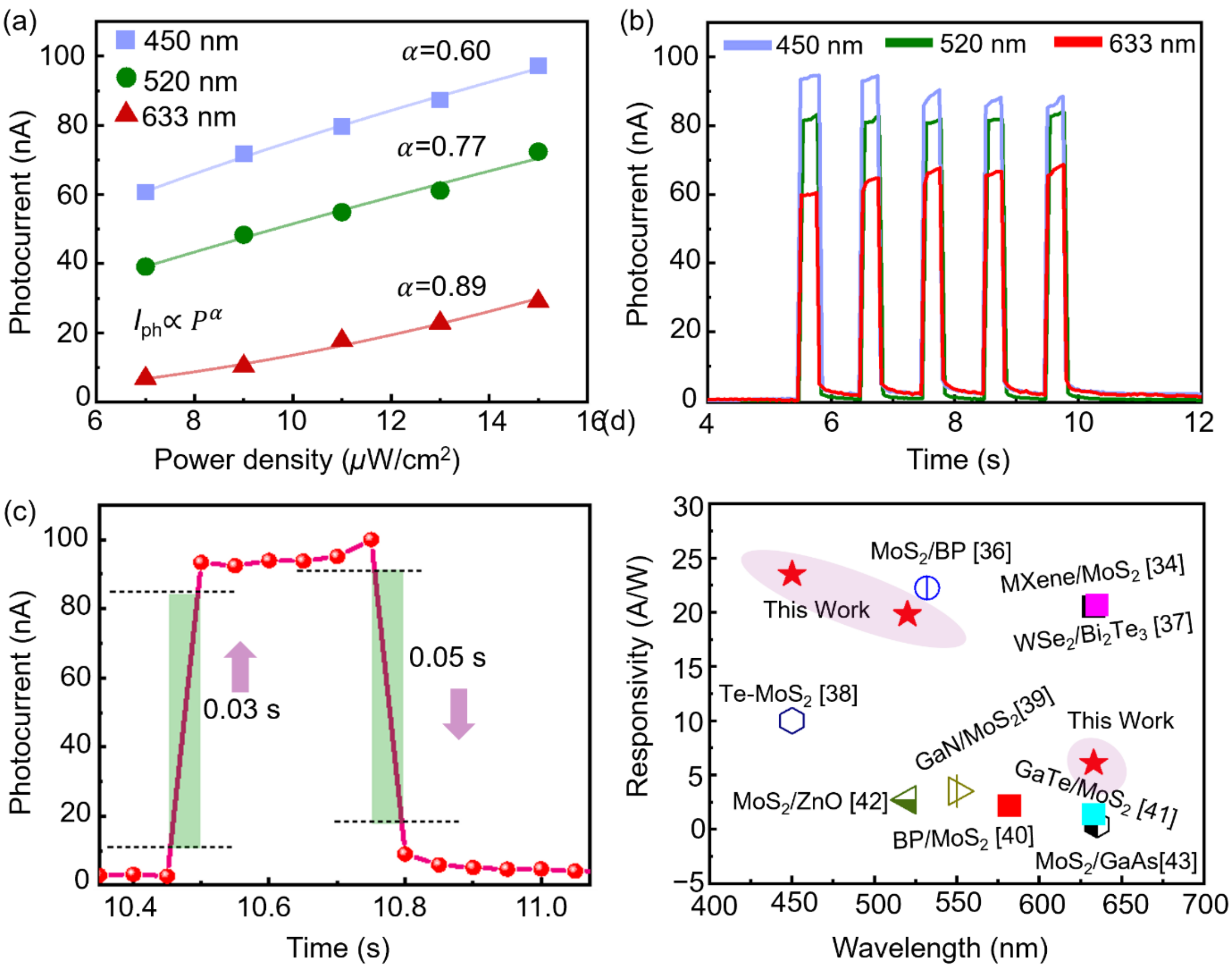

Fig. 4 (a) Power law fitting for the plot of photocurrent versus incident light power density at biasing $V_{ds}$= −1 V and $V_g$= 0 V at different wavelength lasers. (b) Long-cycle stability measurement of the HS device under different wavelength laser illuminations. (c) Corresponding transient photocurrent under 450 nm illumination and $V_{ds}$=−1 V. The pink arrows show the direction of rise and fall of the photocurrent with time. (d) Responsivity comparison of the $WS_2$/FGT Schottky barrier photodetector ($V_{ds}$ = 0 V) against previously reported 2D material-based photodetectors.

The transient response was measured under 450 nm illumination at $V_{ds}$ = −1 V. As shown in Figure 4c, the device shows a clear and repeatable photocurrent response, with rise and fall times of 30 and 50 ms, respectively, extracted between 10% and 90% of the steady-state photocurrent. These response times are consistent with efficient carrier separation and extraction through the HS, while the remaining temporal response may involve trap-mediated carrier capture and release in multilayer $WS_2$. This interpretation is consistent with sublinear power dependence and supports a possible trap-assisted contribution to the photoresponse. The performance of the $WS_2$/FGT HS was benchmarked against representative 2D material-based photodetectors, as summarized in Figure 4d and Table S2. Under zero-bias operation, the present device reaches a responsivity of 23.5 A/W under 450 nm illumination, placing it among competitive self-powered 2D photodetectors.

To elucidate the interfacial origin of the self-powered and high gain photoresponse, density functional theory (DFT) calculations and kelvin probe force microscopy (KPFM) measurements were used to examine the electronic structure and band alignment of the $WS_2$/FGT HS. Figure 5a shows the calculated band structure and density of states of multilayer $WS_2$, confirming its indirect semiconducting nature with a bandgap of approximately 1.35 eV. The valence-band maximum is located near the Γ point, while the conduction-band minimum lies near the K point, consistent with the visible-range optical absorption of multilayer $WS_2$ [11,26]. In contrast, FGT exhibits metallic electronic behavior, with spin-polarized bands crossing the Fermi level (Figure 5b). The corresponding density of states shows a clear asymmetry between spin-up and spin-down channels, confirming the ferromagnetic metallic character of FGT [24,25]. The projected density of states indicates that the electronic and magnetic properties of FGT are mainly governed by Fe 3d orbitals, as shown in Supporting Information Figure S8. This metallicity is essential for device operation because FGT acts as an efficient vdW carrier-collecting electrode at the $WS_2$ interface.

The interfacial energy alignment was evaluated using KPFM measurements and DFT-calculated work functions. Experimentally, KPFM gives work functions of 4.557 eV for $WS_2$ and 4.497 eV for FGT, corresponding to a relative offset of approximately 0.06 eV, as summarized in Supporting Information Figure S5. DFT calculations show the same relative trend (Figure S8), with $WS_2$ having a higher work function than FGT, although the absolute values differ because of surface adsorbates, defects, interface dipoles, and idealized calculation conditions. Based on these results, the band diagram in Figure 5c illustrates the formation of an asymmetric $WS_2$/FGT metal–semiconductor junction. After contact, electrons transfer from lower-work-function FGT toward $WS_2$ until Fermi-level equilibration is reached, producing band bending in $WS_2$ and establishing a built-in electric field across the interface [17,30,31]. Under illumination, this field separates photogenerated electron–hole pairs and enables carrier collection even at zero external bias [34,35].

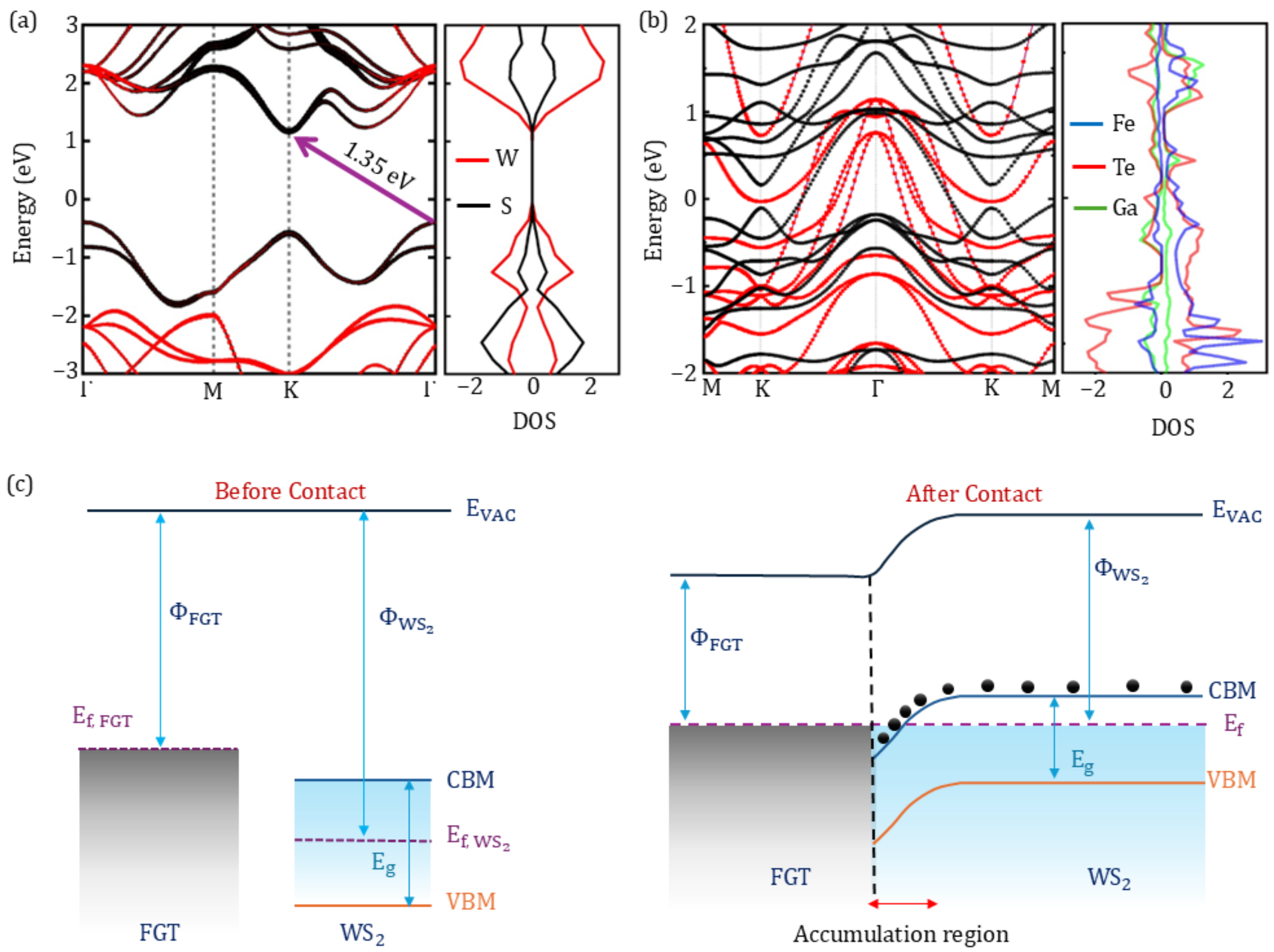

Fig. 5 (a) Calculated band structure and density of states of multilayer $WS_2$, showing an indirect bandgap of approximately 1.35 eV. The projected states indicate contributions from W and S orbitals near the band edges. (b) Spin-polarized band structure and projected density of states of FGT, showing metallic bands crossing the Fermi level and confirming the metallic character of FGT. The orbital-resolved DOS highlights the dominant contribution of Fe, together with Te and Ga states. (c) Schematic energy-band alignment before and after contact between FGT and $WS_2$.

Although the KPFM-derived work-function difference (~60 meV) is more than twice the thermal energy at room temperature ( $k_B T \approx 26$ meV), this comparison does not determine the magnitude of the internal electric field. Therefore, the high zero-bias photoresponse is attributed to the combined effects of interfacial band bending, possible interface dipoles, asymmetric Au/$WS_2$/FGT contact geometry, efficient carrier extraction through metallic FGT, and a possible contribution from photogating-assisted carrier-lifetime extension in multilayer $WS_2$ [32,33]. Together, the DFT and KPFM results support the relative work-function alignment and direction of interfacial charge redistribution at the $WS_2$/FGT interface. Combined with the rectifying transport characteristics and finite zero-bias photocurrent, these results are consistent with an internal electric field that assists photocarrier separation. Metallic FGT facilitates carrier extraction, while trap states in multilayer $WS_2$ provide photogating-assisted gain. These coupled effects explain the high responsivity, large EQE, and zero-bias photoresponse of the $WS_2$/FGT photodetector.

## 3.Conclusions

In summary, we demonstrate a self-powered $WS_2$/FGT vdW metal–semiconductor photodetector with high responsivity and large apparent EQE under weak visible illumination. Under 450 nm illumination, the device achieves a zero-bias responsivity of 23.5 A/W and an apparent EQE of $6.4 \times 10^3$%. At $V_{ds}= -1$ V, it reaches a responsivity of $9.7 \times 10^3$ A/W and a noise-derived specific detectivity of $2.3 \times 10^{13}$ Jones at 100 Hz. The enhanced performance is attributed to interfacial carrier separation and efficient carrier extraction through metallic FGT, together with a likely contribution from trap-assisted photogating in multilayer $WS_2$. KPFM and DFT analyses further support favorable relative work-function alignment and interfacial charge redistribution between FGT and $WS_2$. These results establish FGT-enabled interface engineering as an effective strategy for realizing high-sensitivity, low-power photodetectors based on two-dimensional van der Waals heterostructures.

## Associated Content

**Supporting Information.** Additional structural characterization of FGT and $WS_2$ crystals; AFM, SEM, and EDX analysis of the $WS_2$/FGT heterostructure; PL characterization; device reproducibility; KPFM work-function analysis; noise spectral density measurements; additional DFT details; experimental methods; and theoretical calculation details.

## Author Information

### Corresponding Authors

**Ziwei Li –** Hunan Institute of Optoelectronic Integration, College of Materials Science and Engineering, Hunan University, Changsha, P. R. China.

Email: ziwei_li@hnu.edu.cn

**Maciej R. Molas –** University of Warsaw, Faculty of Physics, 02-093 Warsaw, Poland
Email: maciej.molas@fuw.edu.pl

**Authors**

**Wajid Ali –** University of Warsaw, Faculty of Physics, 02-093 Warsaw, Poland.

**Ming Huang–** Hunan Institute of Optoelectronic Integration, College of Materials Science and Engineering, Hunan University, Changsha, P. R. China.

**Juan Li–** Hunan Institute of Optoelectronic Integration, College of Materials Science and Engineering, Hunan University, Changsha, P. R. China.

**Jianhua Huang–** Hunan Institute of Optoelectronic Integration, College of Materials Science and Engineering, Hunan University, Changsha, P. R. China.

**Liuli Yang–** Hunan Institute of Optoelectronic Integration, College of Materials Science and Engineering, Hunan University, Changsha, P. R. China.

**Sajid Ur Rehman–**School of Science, Minzu University of China, Beijing, P. R. China

**Chinmay K. Mohanty–** University of Warsaw, Faculty of Physics, 02-093 Warsaw, Poland.

**Zahir Muhammad –** National Key Laboratory of Spintronics, Hangzhou International Innovation Institute, Beihang University, Hangzhou 311115, P.R. China.

**Author contributions**

Z.L. and M.R.M. planned and supervised the project. W.A. and J.L. conducted experiments regarding the material synthesis, and characterization of the devices. M.H. and L.Y. supported to building optical setup for Raman and PL measurements. W.A. and C.M. contributed to data analysis and figure preparation. S.U.R. and Z.M. conducted the measurement of thermomagnetic curve and DFT calculations, respectively. W.A., Z.L. and M.R.M. wrote the manuscript with input from all co-authors.
† These authors contributed equally.

**Competing Interest**

The authors declare no competing financial interests.

**Funding**

This work is supported by the European Union's Marie Skłodowska-Curie Actions under the Horizon Europe research and innovation programme (Grant No. 101204286). M.R.Molas gratefully acknowledges support from the National Science Centre, Poland (Grant No. 2022/46/E/ST3/00166 and 2023/50/O/ST3/00310). The work is also supported by the National Key Research and Development Program of China (Grant No. 2022YFA1203801), the National Natural Science Foundation of China (Grant No. 62275076), and the Hunan Provincial Science and Technology Department (Grant Nos. 2026JJ60487, 2023RC3093).

**References**

(1) Eng, I. P. C.; Song, S.; Ping, B. State-of-the-Art Photodetectors for Optoelectronic Integration at Telecommunication Wavelength. J. Mater. Chem. C 2015, 4, 277–302.
(2) Qiu, M.; Sun, P.; Liu, Y.; et al. Visualized UV Photodetectors Based on Prussian Blue/TiO2 for Smart Irradiation Monitoring Application. Adv. Mater. Technol. 2018, 3, 1700288.
(3) Bartolo-Perez, C.; Chandiparsi, S.; Mayet, A. S.; et al. Avalanche Photodetectors with Photon Trapping Structures for Biomedical Imaging Applications. Opt. Express 2021, 29, 19024–19033.
(4) Li, Z.; Yan, T.; Fang, X. Low-Dimensional Wide-Bandgap Semiconductors for UV Photodetectors. Nat. Rev. Mater. 2023, 8, 587–603.
(5) García de Arquer, F. P.; Armin, A.; Meredith, P.; et al. Solution-Processed Semiconductors for Next-Generation Photodetectors. Nat. Rev. Mater. 2017, 2, 16100. Z. Chen, Z. Cheng, J. Wang, X. Wan, C. Shu, H. K. Tsang, H. P. Ho, J. B. Xu, *Adv. Opt. Mater.* **2015**, 3, 1207..
(6) Bai, P.; Li, X.; Yang, N.; Chu, W.; Bai, X.; Huang, S.; Zhang, Y.; Shen, W.; Fu, Z.; Shao, D.; Tan, Z. Broadband and Photovoltaic THz/IR Response in the GaAs-Based Ratchet Photodetector. Sci. Adv. 2022, 8, eabn2031.
(7) Rogalski, A.; Ciupa, R. Performance Limitation of Short Wavelength Infrared InGaAs and HgCdTe Photodiodes. J. Electron. Mater. 1999, 28, 630–636.
(8) Kaniewski, J.; Muszalski, J.; Piotrowski, J. Recent Advances in InGaAs Detector Technology. Phys. Status Solidi A 2004, 20, 2281–2287.
(9) Huo, N.; Konstantatos, G. Recent Progress and Future Prospects of 2D-Based Photodetectors. Adv. Mater. 2018, 30, 1801164.
(10) He, T.; Lan, C.; Zhou, S.; Li, Y.; Yin, Y.; Li, C.; Liu, Y. Enhanced Responsivity of a Graphene/Si-Based Heterostructure Broadband Photodetector by Introducing a WS2 Interfacial Layer. J. Mater. Chem. C 2021, 9, 3846–3853.

(11) Kim, H. C.; Kim, H.; Lee, J. U.; Lee, H. B.; Choi, D. H.; Lee, J. H.; Lee, W. H.; Jhang, S. H.; Park, B. H.; Cheong, H.; Lee, S. W. Engineering Optical and Electronic Properties of WS2 by Varying the Number of Layers. ACS Nano 2015, 9, 6854–6860.
(12) Qu, T.; Fan, J.; Wei, X. Dark Current Reduction and Performance Improvements in Graphene/Silicon HS Photodetectors Obtained Using a Non-Stoichiometric HfOx Thin Oxide Layer. Nanomaterials 2024, 14, 419.
(13) Li, X.; Zhu, M.; Du, M.; Lv, Z.; Zhang, L.; Li, Y.; Yang, Y.; Yang, T.; Li, X.; Wang, K.; Zhu, H. High Detectivity Graphene-Silicon HS Photodetector. Small 2016, 12, 595–601.
(14) Higashitarumizu, N.; Wang, S.; Wang, S.; Kim, H.; Bullock, J.; Javey, A. Black Phosphorus for Mid-Infrared Optoelectronics: Photophysics, Scalable Processing, and Device Applications. Nano Lett. 2024, 24, 13107–13117.
(15) Margot, F.; Lisi, S.; Cucchi, I.; Cappelli, E.; Hunter, A.; Gutiérrez-Lezama, I.; Ma, K.; von Rohr, F.; Berthod, C.; Petocchi, F.; Poncé, S. Electronic Structure of Few-Layer Black Phosphorus from μ-ARPES. Nano Lett. 2023, 23, 6433–6439.
(16) Huang, H.; Sheng, Y.; Zhou, Y.; Zhang, Q.; Hou, L.; Chen, T.; Chang, R. J.; Warner, J. H. 2D-Layer-Dependent Behavior in Lateral Au/WS2/Graphene Photodiode Devices with Optical Modulation of Schottky Barriers. ACS Appl. Nano Mater. 2018, 1, 6874–6881.
(17) Ezhilmaran, B.; Patra, A.; Benny, S.; Mr, S.; Vv, A.; Bhat, S. V.; Rout, C. S. Recent Developments in the Photodetector Applications of Schottky Diodes Based on 2D Materials. J. Mater. Chem. C 2021, 9, 6122–6150.
(18) Kim, J.; Venkatesan, A.; Phan, N. A. N.; Kim, Y.; Kim, H.; Whang, D.; Kim, G. H. Schottky Diode with Asymmetric Metal Contacts on WS2. Adv. Electron. Mater. 2022, 8, 2100941.
(19) Wang, X. F.; Zhao, H. M.; Shen, S. H.; Pang, Y.; Shao, P. Z.; Li, Y. T.; Deng, N. Q.; Li, Y. X.; Yang, Y.; Ren, T. L. High Performance Photodetector Based on Pd-Single Layer MoS2 Schottky Junction. Appl. Phys. Lett. 2016, 109, 201904.
(20) Das, S.; Chen, H. Y.; Penumatcha, A. V.; Appenzeller, J. High Performance Multilayer MoS2 Transistors with Scandium Contacts. Nano Lett. 2013, 13, 100–105.
(21) Li, C.; Wu, Z.; Zhang, C.; Peng, S.; Han, J.; He, M.; Dong, X.; Gou, J.; Wang, J.; Jiang, Y. Self-Powered Photodetector with High Performance Based on All-2D NbSe2/MoSe2 van der Waals Heterostructure. Adv. Opt. Mater. 2023, 11, 2300905.G. Zhang, F. Guo, H. Wu, X. Wen, L. Yang, W. Jin, W. Zhang, and H. Chang, *Nat. Commun*. **2022,** 13, 5067.
(22) Yadav, P.; Bhattacharya, K.; Ganguli, A. K. TaS2 Nanosheets Decorated with SnS2 Nanosheet/Reduced Graphene Oxide Composites for Broadband Photodetectors. ACS Appl. Nano Mater. 2023, 7, 550–559.

(23) Liu, X.; Choi, M. S.; Hwang, E.; Yoo, W. J.; Sun, J. Fermi Level Pinning Dependent 2D Semiconductor Devices: Challenges and Prospects. Adv. Mater. 2022, 34, 2108425.
(24) Wu, S.; He, Z.; Gu, M.; Ren, L.; Li, J.; Deng, B.; Wang, D.; Sou, I. K.; Li, S. Robust ferromagnetism in wafer-scale $Fe_3GaTe_2$ above room-temperature. Nat Commun 2024,15, 10765.
(25) Gao, X.; Zhai, K.; Fu, H.; Yan, J.; Yue, D.; Ke, F.; Zhao, Y.; Mu, C.; Nie, A.; Xiang, J.; Wen, F. Enhanced Ferromagnetism and Tunable Magnetic Anisotropy in a van der Waals Ferromagnet. Adv. Sci. 2024, 11, 2402819.
(26) Berkdemir, A.; Gutiérrez, H. R.; Botello-Méndez, A. R.; Perea-López, N.; Elías, A. L.; Chia, C. I.; Wang, B.; Crespi, V. H.; López-Urías, F.; Charlier, J. C.; Terrones, H. Identification of Individual and Few Layers of WS2 Using Raman Spectroscopy. Sci. Rep. 2013, 3, 1755.
(27) Gao, W.; Zhang, S.; Zhang, F.; Wen, P.; Zhang, L.; Sun, Y.; Chen, H.; Zheng, Z.; Yang, M.; Luo, D.; Huo, N. 2D WS2 Based Asymmetric Schottky Photodetector with High Performance. Adv. Electron. Mater. 2021, 7, 2000964.
(28) Luo, Z.; Xu, H.; Gao, W.; Yang, M.; He, Y.; Huang, Z.; Yao, J.; Zhang, M.; Dong, H.; Zhao, Y.; Zheng, Z. High-Performance and Polarization-Sensitive Imaging Photodetector Based on WS2/Te Tunneling Heterostructure. Small 2023, 19, 2207615.
(29) Duan, J.; Chava, P.; Ghorbani-Asl, M.; Lu, Y.; Erb, D.; Hu, L.; Echresh, A.; Rebohle, L.; Erbe, A.; Krasheninnikov, A. V.; Helm, M. Self-Driven Broadband Photodetectors Based on MoSe2/FePS3 van der Waals n–p Type-II Heterostructures. ACS Appl. Mater. Interfaces 2022, 14, 11927–11936.
(30) Pak, Y.; Park, W.; Alaal, N.; Kumaresan, Y.; Aravindh, S. A.; Mitra, S.; Xin, B.; Min, J. W.; Kim, H.; Lim, N.; Cho, B. Enhanced Photoresponse of WS2 Photodetectors Through Interfacial Defect Engineering Using a TiO2 Interlayer. ACS Appl. Electron. Mater. 2020, 2, 838–845.
(31) Kim, S.; Kim, M.; Kim, H. Self-Powered Photodetectors Based on Two-Dimensional van der Waals Semiconductors. Nano Energy 2024, 127, 109725.
(32) Zou, J.; Huang, Y.; Wang, W.; Li, C.; Wei, S.; Liu, H.; Luo, L.; Du, W.; Shen, K.; Ren, A.; Wu, J. Plasmonic MXene Nanoparticle-Enabled High-Performance Two-Dimensional MoS2 Photodetectors. ACS Appl. Mater. Interfaces 2022, 14, 8243–8250.
(33) Hu, X.; Hou, J.; Dong, Q.; Liu, Z.; Wang, S.; Wang, F.; Tu, Y.; Han, T.; Li, F.; Zhang, Z.; Hou, X. Ultrabroadband Photodetector Based on Ferromagnetic van der Waals Heterodiode. Adv. Electron. Mater. 2022, 8, 2200208.
(34) Ye, L.; Li, H.; Chen, Z.; Xu, J. Near-Infrared Photodetector Based on MoS2/Black Phosphorus HS. ACS Photonics 2016, 3, 692–699.

(35) Pecunia, V.; Anthopoulos, T. D.; Armin, A.; Bouthinon, B.; Caironi, M.; Castellanos-Gomez, A.; Chen, Y.; Cho, K.; Clegg, C.; Fang, X.; Fendel, P.; Fowler, B.; Gelinck, G.; Gottlob, H.; Guyot-Sionnest, P.; Hannebauer, R.; Hernandez-Sosa, G.; Hersam, M. C.; Hirsch, L.; Ho, J. C.; Isikgor, F. H.; Joimel, J.; Kim, H. J.; Konstantatos, G.; Labram, J.; Lemme, M. C.; Leo, K.; Lhuillier, E.; Lidorikis, E.; Loi, M. A.; Malinowski, P. E.; Merken, P.; Mueller, T.; Nasrollahi, B.; Natali, D.; Ng, T. N.; Nguyen, T.-Q.; Park, S. K.; Peng, L.-M.; Samorì, P.; Sargent, E. H.; Shen, L.; Shishido, S.; Shorubalko, I.; Sonar, P.; Stranks, S. D.; Tedde, S. F.; Vandewal, K.; Verhaegen, M.; Walia, S.; Yan, F.; Yokota, T.; Zhang, F. Guidelines for Accurate Evaluation of Photodetectors Based on Emerging Semiconductor Technologies. Nat. Photonics 2025, 19, 1178–1188.

(36) Park, J.; Kim, Y.; You, B.; Huh, J.; Kim, G.; Son, H.; Park, Y.; Hahm, M. G.; Kim, U. J.; Lee, M. High-Performance 1D–2D Te/MoS2 Heterostructure Photodetectors with Tunable Giant Persistent Photoconductivity. ACS Appl. Electron. Mater. 2024, 6, 6147–6154.

(37) Janardhanam, V.; Zummukhozol, M.; Jyothi, I.; Shim, K. H.; Choi, C. J. Self-Powered MoS2/n-Type GaN Heterojunction Photodetector with Broad Spectral Response in Ultraviolet–Visible–Near-Infrared Range. Sens. Actuators, A 2023, 360, 114534.

(38) Zheng, S.; Wu, E.; Feng, Z.; Zhang, R.; Xie, Y.; Yu, Y.; Zhang, R.; Li, Q.; Liu, J.; Pang, W.; Zhang, H. Acoustically Enhanced Photodetection by a Black Phosphorus–MoS2 van der Waals Heterojunction p–n Diode. Nanoscale 2018, 10, 10148–10153.S. Yang, C. Wang, C. Ataca, Y. Li, H. Chen, H. Cai, A. Suslu, J. C. Grossman, C. Jiang, Q. Liu, and S. Tongay, *ACS Appl. Mater. & Interfaces* 2016, 8, 2533-2539.

(39) Yang, S.; Wang, C.; Ataca, C.; Li, Y.; Chen, H.; Cai, H.; Suslu, A.; Grossman, J. C.; Jiang, C.; Liu, Q.; Tongay, S. Self-Driven Photodetector and Ambipolar Transistor in Atomically Thin GaTe-MoS2 p–n vdW Heterostructure. ACS Appl. Mater. Interfaces 2016, 8, 2533–2539.

(40) Kang, M. A.; Kim, S.; Jeon, I. S.; Lim, Y. R.; Park, C. Y.; Song, W.; Lee, S. S.; Lim, J.; An, K. S.; Myung, S. Highly Efficient and Flexible Photodetector Based on MoS2–ZnO Heterostructures. RSC Adv. 2019, 9, 19707–19711.

(41) Xu, Z.; Lin, S.; Li, X.; Zhang, S.; Wu, Z.; Xu, W.; Lu, Y.; Xu, S. Monolayer MoS2/GaAs Heterostructure Self-Driven Photodetector with Extremely High Detectivity. Nano Energy 2016, 23, 89–96.

(42) Lee, S. H.; Lee, D.; Hwang, W. S.; Hwang, E.; Jena, D.; Yoo, W. J. High-Performance Photocurrent Generation from Two-Dimensional WS2 Field-Effect Transistors. Appl. Phys. Lett. 2014, 104, 193113.

(43) Kim, J.; Venkatesan, A.; Phan, N. A. N.; Kim, Y.; Kim, H.; Whang, D.; Kim, G. H. Schottky Diode with Asymmetric Metal Contacts on $WS_2$. Adv. Electron. Mater. 2022, 0, 2100941.

# Supporting Information

## S1. Growth and Characterization of FGT

High-purity, gallium (Ga) foil (99.99%, Sigma-Aldrich), iron (Fe) powder (99.98%), and tellurium (Te) powder (99.98%, Alfa Aesar) were used as precursors for $Fe_3GaTe_2$ crystal growth. FGT single crystals were synthesized using a self-flux method. High-purity Fe, Ga, and Te were mixed in a 1:1:2 ratio, sealed in an evacuated quartz tube, heated to 1000 °C for 3 hours, held for 24 hours, and then slowly cooled to 770 °C over 100 hours. Figure S1a shows the X-ray diffraction (XRD) pattern of the FGT bulk crystal, oriented along the (00l) direction. This orientation indicates that the crystal's surface is perpendicular to the c-axis, aligning the plate-like surface with the ab plane. The distinct and intense diffraction peaks signify the high quality of the directly grown samples. Advanced lattice parameter refinement and crystal structure elucidation were achieved through single crystal XRD analysis. Figure S1b illustrates the energy-dispersive X-ray (EDX) spectrum, which was utilized to ascertain the elemental ratios within the crystal. The elemental composition was deduced from the integrated intensity ratios of the distinct peaks corresponding to each element. The average of these peak intensities across three different types of peaks provided a composition estimate that closely aligns with the target composition for FGT. Figure 1c and 1d shows the isothermal magnetization M(H) curves measured at 300 K and 200 K, respectively. These curves clearly exhibit the characteristic ferromagnetic ordering behavior of FGT, as evidenced by the well-defined hysteresis loops. The data confirm the robust ferromagnetic nature of FGT at these temperatures, consistent with its known magnetic properties. Furthermore, these results are in excellent agreement with previously reported studies, reinforcing the reliability of the measurements and the intrinsic ferromagnetic behavior of FGT.

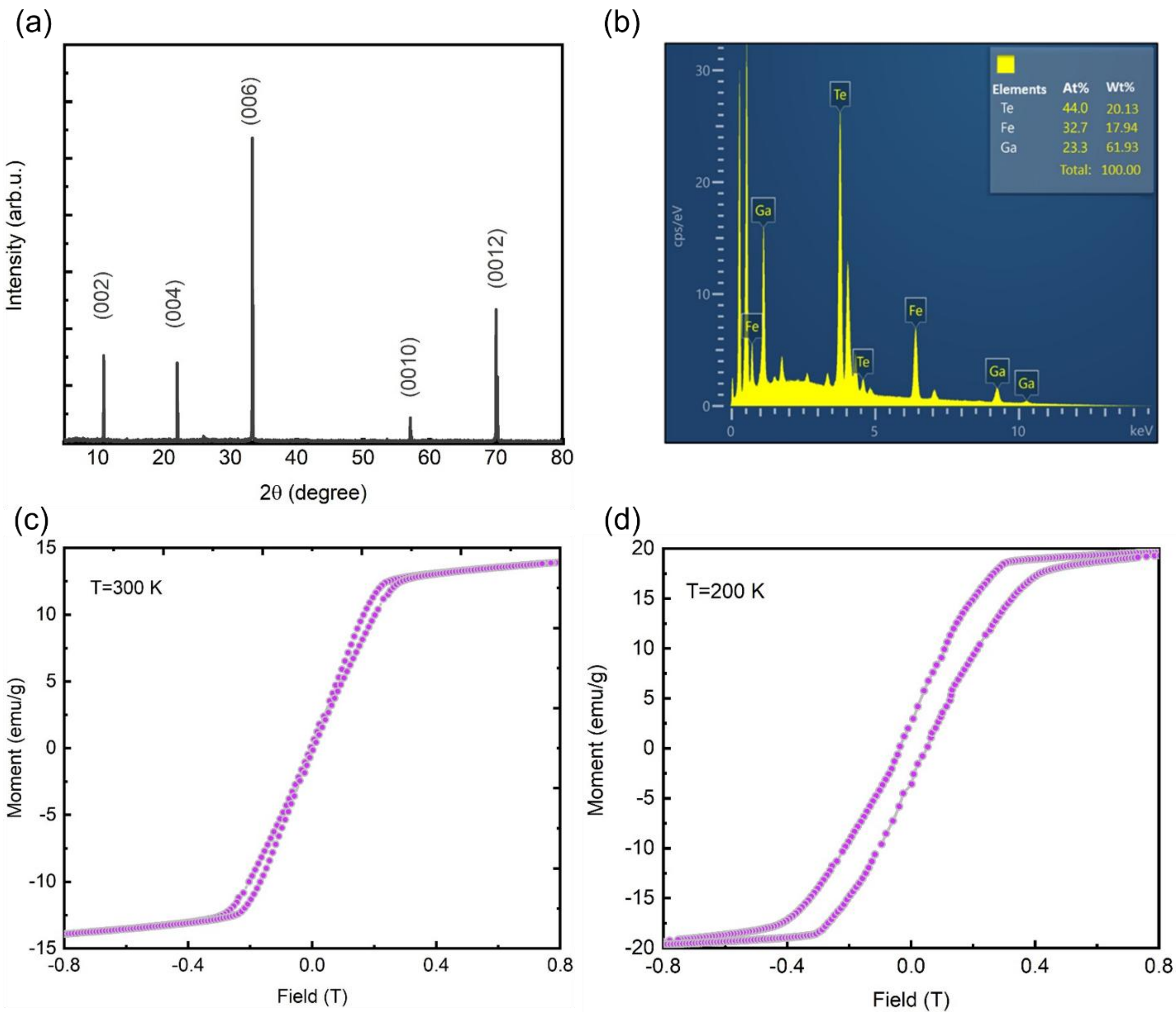

Fig.S1 (a) FGT XRD spectrum at room temperature; (b) EDX spectrum shows elemental analysis of FGT. (c, d) The isothermal magnetization M(H) of FGT at 300 K and 200 K respectively.

## S2. AFM analysis of $WS_2$/FGT

Surface morphology and thickness of the $WS_2$/FGT heterostructure were characterized using a Bruker Dimension Icon atomic force microscope operating in tapping mode. The corresponding height profile of $WS_2$ and FGT are shown in Figure S2(a, b).

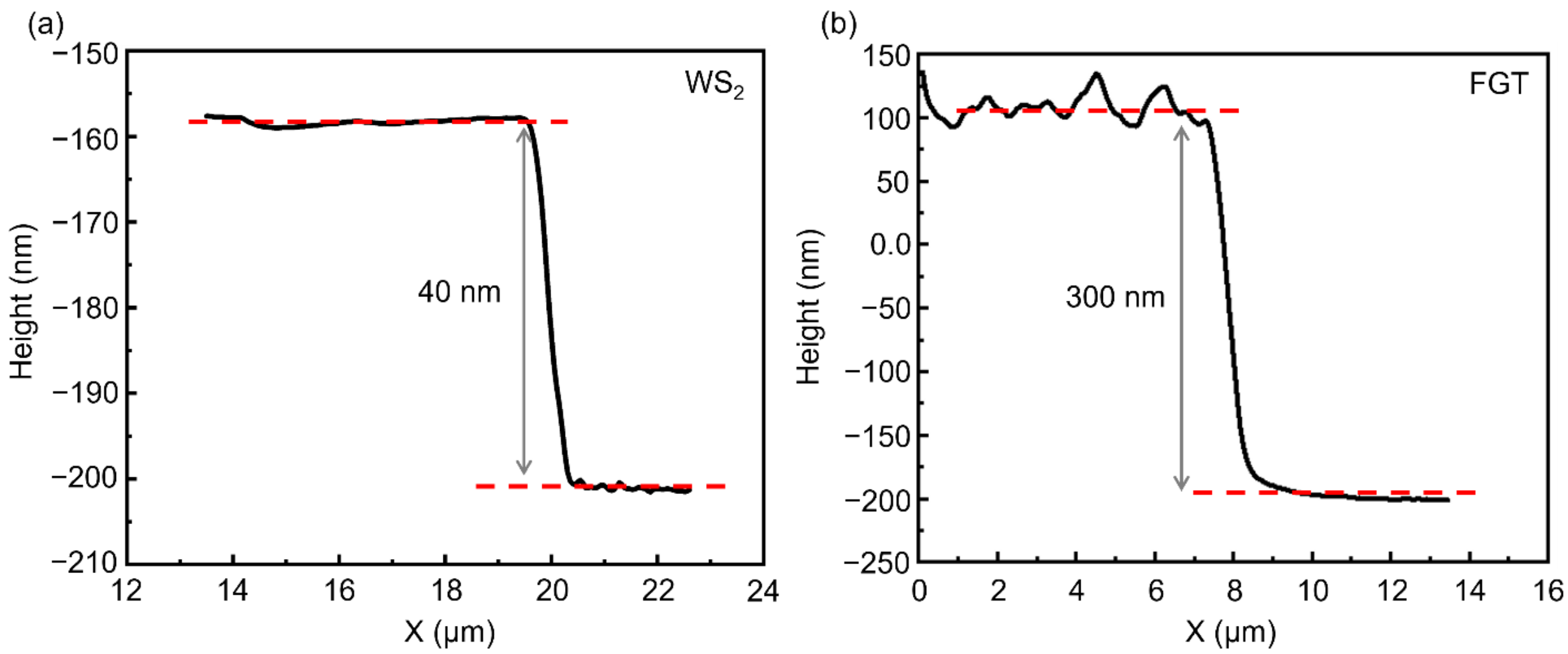

Fig. S2 Thickness determination by Atomic Force Microscopy (AFM). AFM height profiles show thicknesses of (a) 40 nm for $WS_2$ and (b) 300 nm for FGT.

## S3. Photoluminescence of $WS_2$/FGT heterojunction

The $WS_2$/FGT heterostructure devices were fabricated using mechanical exfoliation and dry-transfer technique. Multilayer $WS_2$ and FGT flakes were exfoliated using adhesive tape and first transferred onto polydimethylsiloxane (PDMS) substrates. Suitable flakes were identified using optical microscopy prior to transfer. The FGT flake was first transferred onto a pre-patterned Au electrode fabricated by standard photolithography followed by electron-beam evaporation. Subsequently, a multilayer $WS_2$ flake was transferred such that it partially overlapped the FGT flake while extending onto a second Au electrode, forming a $WS_2$/FGT metal–semiconductor HS device.

PL, Raman spectroscopy, and spatial mapping measurements were performed using a WITec alpha-300 confocal microscope equipped with a UHTS300 spectrometer. A 532 nm excitation laser was used together with a 50× objective lens (Zeiss, NA = 0.75), providing an effective laser spot size of approximately 1 µm. The laser power was maintained at 0.25 mW during measurements. An 1800 grooves $mm^{-1}$ grating was used for Raman spectral acquisition. The noise current spectra were measured using a KEYSIGHT 35670A dynamic signal analyzer. To reduce external electromagnetic interference, the device was enclosed inside a shielded metal measurement chamber during noise characterization.

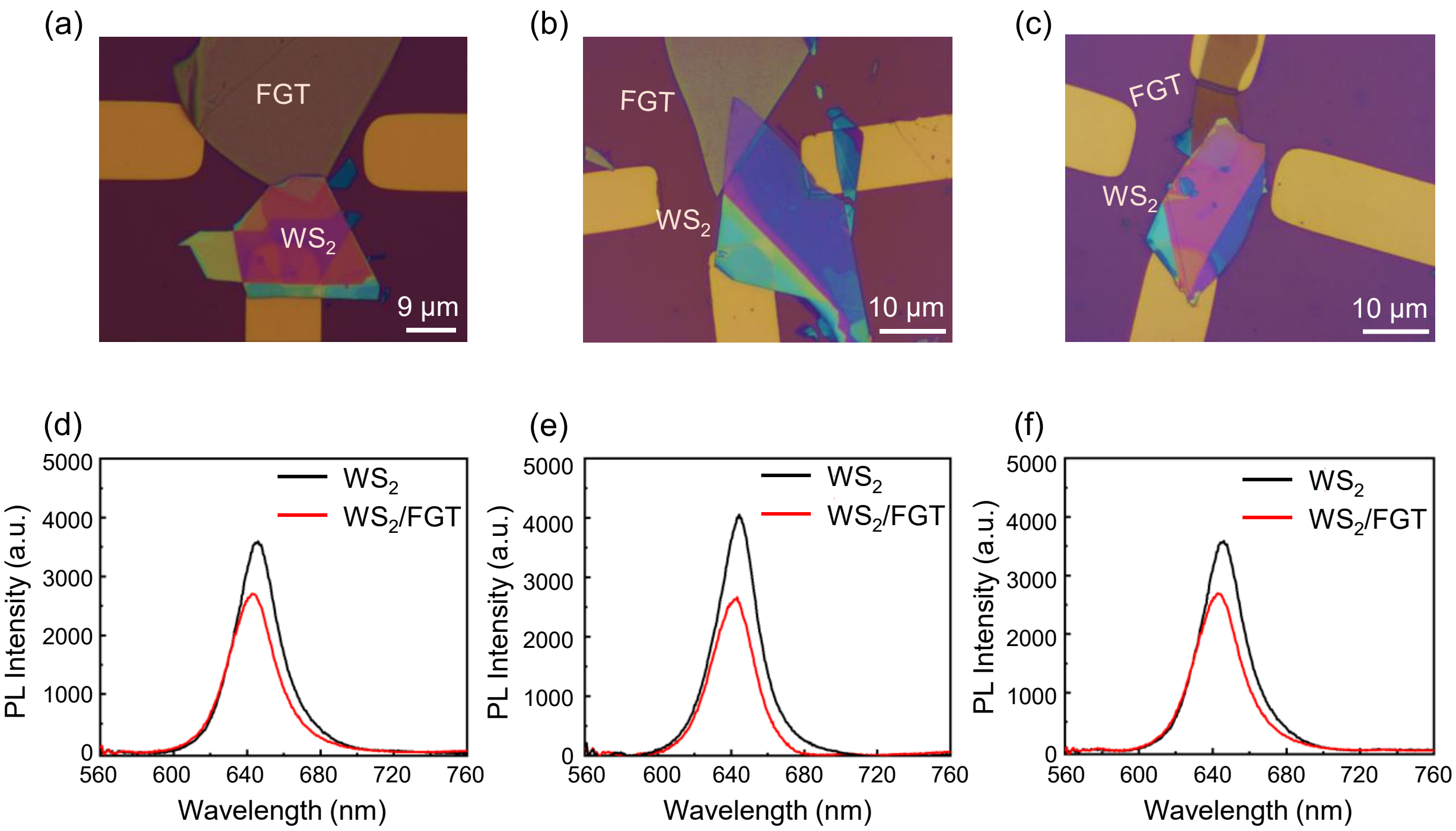

Fig. S3 Optical and PL characterization of $WS_2$/FGT heterostructures. (a–c) Optical microscopy images of three distinct $WS_2$/FGT samples. (d–f) Corresponding PL spectra obtained from the $WS_2$ region and the $WS_2$/FGT heterostructure region in each sample. The reduced PL intensity in the $WS_2$/FGT region indicates PL quenching, suggesting interfacial coupling between $WS_2$ and FGT.

## S4. Comparison of I–V curves of various $WS_2$/FGT devices

Electrical transport measurements of the $WS_2$/FGT heterostructure devices were performed using a CRX-VF probe station equipped with an Agilent B1500 semiconductor parameter analyzer. Prior to measurement, the device chamber was evacuated to reduce the influence of ambient adsorbates on electrical performance. A two-probe configuration was employed, in which the drain–source voltage (Vds) was applied across the $WS_2$/FGT HS device, and the corresponding current (Ids) was recorded. The voltage sweep range was from −1 V to +1 V to evaluate the current–voltage characteristics of the device. No external gate voltage was applied during measurements (Vg = 0 V), allowing characterization of the intrinsic transport behavior of the $WS_2$/FGT HS. All measurements were performed at room temperature.

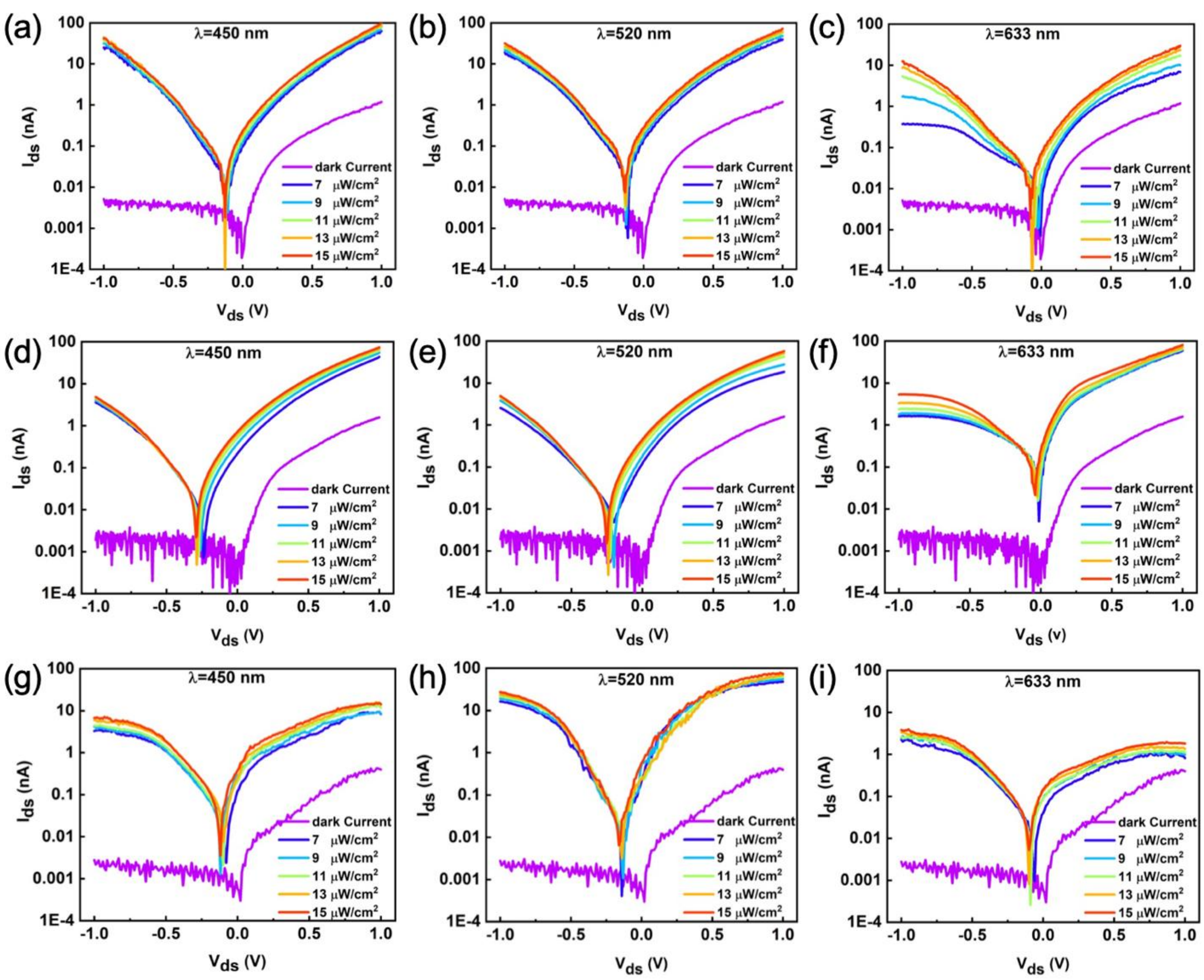

Fig. S4 Photoresponse characteristics of three devices under dark and illuminated conditions with different excitation wavelengths (450 nm, 520 nm, and 633 nm) and power densities (7–15 μW $cm^{-2}$). Panels (a–c) correspond to device 1, (d–f) to device 2, and (g–i) to device 3. Performance metrics of all devices are listed in table S1.

Table S1. Performance metrics of $WS_2$/FGT photodetector at $V_{ds}$=0V

| S. No. | Wavelength (nm) | Responsivity ($AW^{-1}$) | EQE (%) |
|---|---|---|---|
| Device 01 | 450 | 23.5 | $6.4\times10^3$ |
| | 520 | 19.8 | $4.7\times10^3$ |
| | 633 | 6.1 | $1.2\times10^3$ |
| Device 02 | 450 | 20.2 | $5.5\times10^3$ |
| | 520 | 18.2 | $4.3\times10^3$ |
| | 633 | 5.3 | $1.0\times10^3$ |
| Device 03 | 450 | 17.3 | $4.7\times10^3$ |
| | 520 | 14.2 | $3.3\times10^3$ |
| | 633 | 4.8 | $0.9\times10^3$ |

**S5. KPFM image of the $WS_2$/FGT heterostructure**

The surface potential distribution across the heterostructure was investigated using Kelvin probe force microscopy (KPFM) in AC mode with a conductive probe. The contact potential difference (CPD) between the tip and sample surface was measured to evaluate the relative work-function variation across the $WS_2$/FGT interface. To calibrate the tip, a gold film was used as a reference material, giving an electronic surface potential difference of gold (eSPDAu) = − 0.07793 V = − 77.93 mV. The tip work function $\Phi_{tip}$ was calculated as follows:

$$\Phi_{tip} = \Phi_{Au} + eSPDAu = 5.022 \text{ eV}$$

Figure S5a and S5b shows the KPFM image of the $WS_2$/FGT heterostructure and the corresponding surface potential profile. From this we determined the electronic surface potential differences to be $eSPDWS_2$ = 0.465 V and eSPDFGT = 0.525 V. The work functions for $WS_2$ and FGT were then calculated as:

$$\Phi WS_2 = \Phi_{tip} - eSPDWS_2 = 4.557 \text{ eV}$$

$$\Phi FGT = \Phi_{tip} - eSPDFGT = 4.497 \text{ eV}$$

The lower work function of FGT relative to $WS_2$ indicates that electrons tend to transfer from FGT to $WS_2$ until Fermi-level equilibrium is reached, resulting in downward band bending on the $WS_2$ side near the FGT interface and the formation of an interfacial built-in electric field.

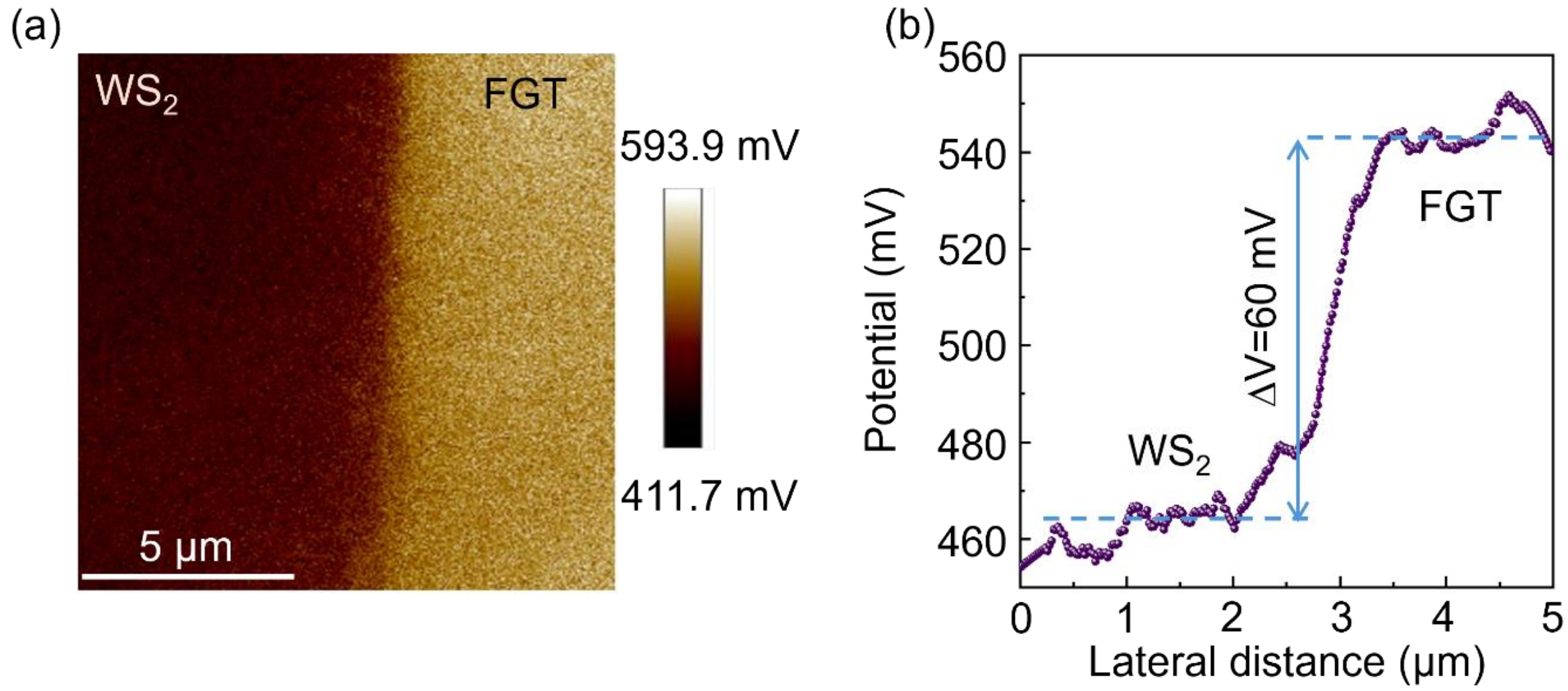

Fig. S5 (a, b) SKPM image and surface potential profile across the $WS_2$/FGT interface, repsectively. A surface potential difference of $\Delta V \approx 60$ mV between the FGT and $WS_2$ is obtained.  The higher surface potential of FGT indicates a lower work function relative to $WS_2$.

## S6. Measured noise spectral density of the $WS_2$/FGT device

The specific detectivity $D^*$ is a crucial figure of merit that quantifies the sensitivity of a photodetector, reflecting its ability to discern weak optical signals from inherent noise. This parameter allows for direct comparisons of device performance across varying active areas and geometries, providing insights into the minimum detectable signal. In our device, the measured current noise spectral density exhibits a characteristic 1/f dependence at low frequencies, indicating dominant flicker-noise contributions (Figure

S6). At a modulation frequency of 100 Hz, the experimental noise current density $I_n$ is $3.75\times10^{-13}$ A Hz $^{-1/2}$. Based on this measured noise, the Noise Equivalent Power (NEP) can be calculated as:

$$\mathrm{NEP}(f) = \frac{I_n\,(f)}{R}$$

where $R$ =9.7 $\times10^{3}$ A/W is the responsivity of the device at $V_{ds}$=−1V. Substituting these values:

$$\mathrm{NEP} = 3.87\times10^{-17}\ \mathrm{W\ Hz^{-1/2}}$$

The specific detectivity $D^*$ is then expressed as:

$$D^*(f) = \frac{\sqrt{A}}{NEP} = \frac{R\sqrt{A}}{I_n(f)}$$

Here, A =80 $\mu$m$^2$ represents the effective area of the photodetector, and $f_b$ denotes the noise frequency bandwidth (1Hz). Substituting the experimental NEP yields:

$$D^* = 2.3\times10^{13}\ \text{Jones}$$

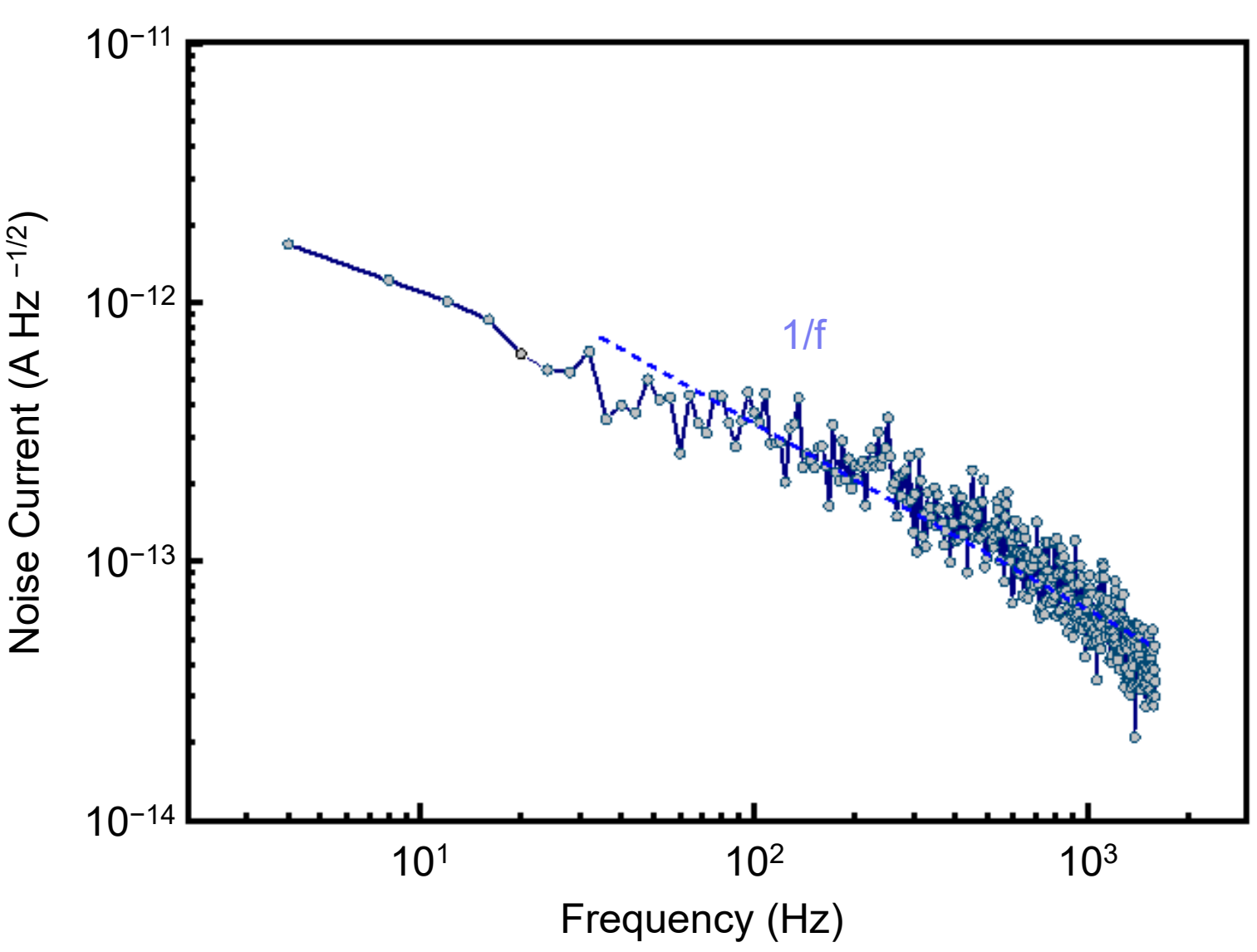

Fig.S6 Measured frequency-dependent noise current spectrum of the $WS_2$/FGT device at $V_{ds}$= −1 V.

**S7. Comparison of photoresponse of the $WS_2$/FGT device with the counterpart devices**

Table S2. Photodetection Performance Comparison of This Work with Similar Devices

| Material | Operation Condition ($V_g$, $V_{ds}$) | Wavelength (nm) | Responsivity (A $W^{-1}$) | Ref. |
|---|---|---|---|---|
| CdSe/Graphene | 0 | 633 | $10 \times 10^{-3}$ | [1] |
| Si/Graphene | 0 | 850 | $730 \times 10^{-3}$ | [2] |
| $MoS_2/WS_2$ | 0 | 532 | $4.36 \times 10^{-3}$ | [3] |
| Graphene/$MoSe_2$/Au | 0 | 450 | $89.5 \times 10^{-3}$ | [4] |
| $BiVO_4$/MXene | 0 | 447 | $790.2 \times 10^{-3}$ | [5] |
| GaTe-$MoS_2$ | 0 | 514 | 1.36 | [6] |
| $MoS_2$/BP | 0 | 532 | 22.3 | [7] |
| GeAs/InSe | 0 | 405 | 0.36 | [8] |
| $MoSe_2/FePS_3$ | 0 | 522 | $52 \times 10^{-3}$ | [9] |
| $WS_2$/FGT | 0 | 450 | 23.5 | This Work |
| $WS_2$/FGT | 0 | 520 | 19.8 | This Work |
| $WS_2$/FGT | 0 | 633 | 6 | This Work |

**S8. DFT calculated work functions of $WS_2$ and FGT**

The electronic structures of pristine $WS_2$, FGT, and the $WS_2$/FGT heterostructure were calculated using density functional theory (DFT) as implemented in the Vienna Ab initio Simulation Package (VASP). Spin-polarized calculations were performed to account for the intrinsic ferromagnetic nature of FGT. A vacuum layer of 20 Å was introduced along the out-of-plane direction to eliminate artificial interactions between periodic slabs. The interlayer van der Waals interaction between $WS_2$ and FGT was described using the Grimme DFT-D3 correction method. The on-site Coulomb interaction of Fe 3d orbitals was treated using the DFT + U approach with an effective Hubbard parameter $U_{eff}$ = 4 eV. The exchange–correlation interaction was described using the Perdew–Burke–Ernzerhof (PBE) functional within the generalized gradient approximation (GGA), and the plane-wave cutoff energy was set to 500 eV. All atomic structures were fully relaxed until the residual forces were less than $10^{-2}$ eV $Å^{-1}$ and the total energy convergence criterion reached $10^{-6}$ eV. The Brillouin zone was sampled using a Γ–centered 11×11×1 Monkhorst–Pack k–point mesh.

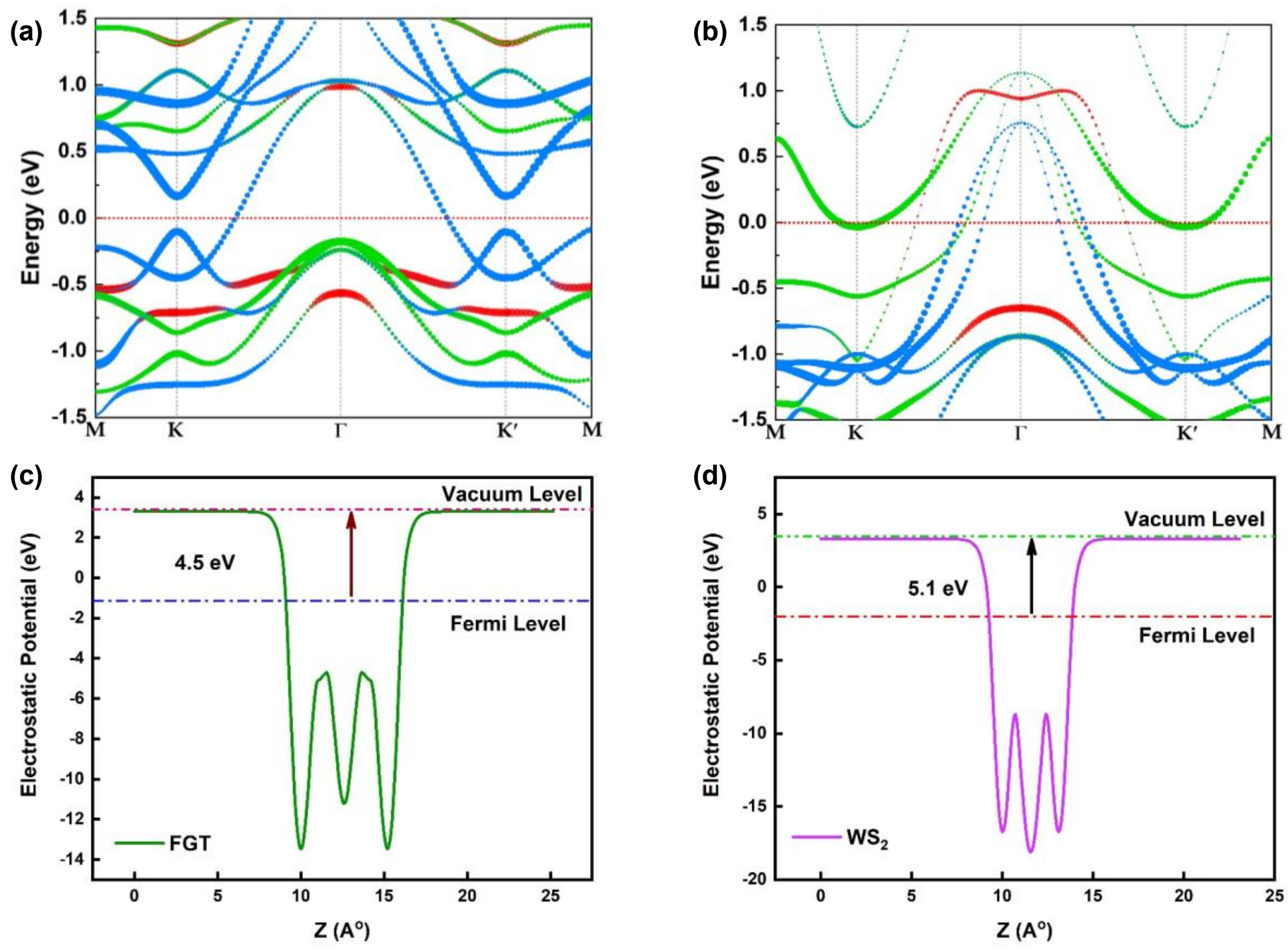

Fig. S8 (a, b) Spin-resolved electronic band structures of ferromagnetic FGT, showing spin-down and spin-up states, respectively. The orbital contributions are indicated by color: blue ($d_{xy}$+$d_{x^2-y^2}$), green ($d_{zx}$+$d_{yz}$) and red ($d_{z^2}$). (c, d) Calculated electrostatic potential and work function diagrams for FGT and multilayer $WS_2$, respectively.

## References

(1) Jin, W.; Ye, Y.; Gan, L.; Yu, B.; Wu, P.; Dai, Y.; Meng, H.; Guo, X.; Dai, L. Self-Powered High Performance Photodetectors Based on CdSe Nanobelt/Graphene Schottky Junctions. J. Mater. Chem. 2012, 22, 2863–2867.

(2) Xiang, D.; Han, C.; Hu, Z.; Lei, B.; Liu, Y.; Wang, L.; Hu, W. P.; Chen, W. Surface Transfer Doping-Induced, High-Performance Graphene/Silicon Schottky Junction-Based, Self-Powered Photodetector. Small 2015, 11, 4829–4836.

(3) Wu, W.; Zhang, Q.; Zhou, X.; Li, L.; Su, J.; Wang, F.; Zhai, T. Self-Powered Photovoltaic Photodetector Established on Lateral Monolayer MoS2-WS2 Heterostructures. Nano Energy 2018, 51, 45–53.

(4) Liu, B.; Zhao, C.; Chen, X.; Zhang, L.; Li, Y.; Yan, H.; Zhang, Y. Self-Powered and Fast Photodetector Based on Graphene/MoSe2/Au Heterojunction. Superlattices Microstruct. 2019, 130, 87–92.

(5) Zhou, S.; Jiang, C.; Han, J.; Mu, Y.; Gong, J. R.; Zhang, J. High-Performance Self-Powered PEC Photodetectors Based on 2D BiVO4/MXene Schottky Junction. Adv. Funct. Mater. 2025, 35, 2416922.

(6) Yang, S.; Wang, C.; Ataca, C.; Li, Y.; Chen, H.; Cai, H.; Suslu, A.; Grossman, J. C.; Jiang, C.; Liu, Q.; Tongay, S. Self-Driven Photodetector and Ambipolar Transistor in Atomically Thin GaTe-MoS2 p–n vdW Heterostructure. ACS Appl. Mater. Interfaces 2016, 8, 2533–2539.

(7) Ye, L.; Li, H.; Chen, Z.; Xu, J. L. Near-Infrared Photodetector Based on $MoS_2$/Black Phosphorus Hetrojunction. ACS Photnics 2016, 3, 692–699.

(8) Xiong, J.; Sun, Y.; Wu, L.; Wang, W.; Gao, W.; Huo, N.; Li, J. High Performance Self-Driven Polarization-Sensitive Photodetectors Based on GeAs/InSe Heterojunction. Adv. Opt. Mater. 2021, 9, 2101017.

(9) Duan, J.; Chava, P.; Ghorbani-Asl, M.; Lu, Y.; Erb, D.; Hu, L.; Echresh, A.; Rebohle, L.; Erbe, A.; Krasheninnikov, A. V.; Helm, M. Self-Driven Broadband Photodetectors Based on MoSe2/FePS3 van der Waals n–p Type-II Heterostructures. ACS Appl. Mater. Interfaces 2022, 14, 11927–11936.